\documentclass[sigconf]{acmart}

\AtBeginDocument{%
  }

\usepackage{pifont}
\usepackage[utf8]{inputenc} 
\usepackage{amsfonts}
\usepackage{xspace}
\usepackage[most]{tcolorbox}
\usepackage{fontawesome5}

\def\eg{\emph{e.g.,}\xspace} 

\def\ie{\emph{i.e.,}\xspace}

\newtcolorbox{boxK}{
    top=2.2pt,
    bottom=2.2pt,
    left=4.5pt,
    right=4.5pt,
    boxrule = 0pt,
    toprule = 0pt, 
    enhanced,
}

\begin{document}

\title{AI-Driven Self-Evolving Software: A Promising Path Toward Software Automation}

\author{Liyi Cai}
\authornote{Both authors contributed equally to this research.}
\affiliation{%
  \institution{College of AI, Tsinghua University}
  \city{Beijing}
  \country{China}
}
\affiliation{%
  \institution{School of Computer Science, Peking University}
  \city{Beijing}
  \country{China}
}
\email{cailiyi@stu.pku.edu.cn}

\author{Yijie Ren}
\authornotemark[1]
\affiliation{%
  \institution{College of AI, Tsinghua University}
  \city{Beijing}
  \country{China}
}
\affiliation{%
  \institution{School of Computer Science and Engineering, Beihang University}
  \city{Beijing}
  \country{China}
}
\email{23373276@buaa.edu.cn}

\author{Yitong Zhang}
\authornotemark[1]
\affiliation{%
  \institution{College of AI, Tsinghua University}
  \city{Beijing}
  \country{China}
}
\email{22373337@buaa.edu.cn}

\author{Jia Li}
\authornote{Jia Li is the corresponding author.}
\affiliation{%
  \institution{College of AI, Tsinghua University}
  \city{Beijing}
  \country{China}
}
\email{jia_li@mail.tsinghua.edu.cn}

\begin{abstract}
Software automation has long been a central goal of software engineering, striving for software development that proceeds without human intervention.
Recent efforts have leveraged Artificial Intelligence (AI) to advance software automation with notable progress. However, current AI functions primarily as assistants to human developers, leaving software development still dependent on explicit human intervention.
This raises a fundamental question: \textit{Can AI move beyond its role as an assistant to become a core component of software, thereby enabling genuine software automation?} 
To investigate this vision, we introduce \textbf{AI-Driven Self-Evolving Software}, a new form of software that evolves continuously through direct interaction with users. 
We demonstrate the feasibility of this idea with a lightweight prototype built on a multi-agent architecture that autonomously interprets user requirements, generates and validates code, and integrates new functionalities. Case studies across multiple representative scenarios show that the prototype can reliably construct and reuse functionality, providing early evidence that such software systems can scale to more sophisticated applications and pave the way toward truly automated software development.
We make code and cases in this work publicly available at {\color{blue}\url{https://github.com/Cai-bird-one/live-software}}.
\end{abstract}

\maketitle

\section{Introduction}

\begin{center}
    \itshape
    Software is eating the world, but AI is going to eat software.
\end{center}
\begin{flushright}
\itshape
    --- Jensen Huang (CEO of NVIDIA)
\end{flushright}
\vspace{3pt}

As software increasingly serves as the foundation of modern society~\cite{lalband2019software, bjorner2003new, vcelebic2023systematic}, researchers and industry practitioners have devoted growing attention to advancing software automation~\cite{dustin1999automated, fu2022vulrepair, awad2024artificial}, whose central goal is to achieve software development without human intervention~\cite{yang2024swe, yarlagadda2021software}.
In this context, one of the most promising directions is the use of AI~\cite{achiam2023gpt}, which leverages its powerful capabilities in understanding, reasoning, and generation to advance software automation across multiple stages of the software lifecycle~\cite{wang2024application, rajendran2025multi, li2025large, schafer2023empirical, shah2024streamlining}.

However, despite considerable progress, the level of automation in current software engineering remains far from ideal. Existing advanced AI in software engineering primarily functions as assistants or just productivity tools for human developers~\cite{nguyen2025generative, durrani2024decade, khemka2024toward}. They can assist in code generation, accelerate testing, and provide design suggestions~\cite{zhang2023survey, applis2025unified, difftester}, but software development still follows the traditional multi-stage lifecycle that requires explicit human intervention at every step. This reliance on human developers unavoidably leads to substantial economic costs~\cite{cui2025effects}, while the handover and coordination across multiple stages inevitably incur considerable time overhead~\cite{saeeda2023challenges}.

Based on the remarkable potential demonstrated by recent advances in AI and its significant achievements in software engineering~\cite{achiam2023gpt, zhang2025beyond, jiang2024survey, li2025aixcoder, jiang2025aixcoder}, we argue that AI may hold the key to realizing genuine software automation. We therefore pose a fundamental question: \textit{Can AI move beyond its role as an assistant to become a core component of software, enabling software to self-evolve without human involvement?}

To explore this possibility, we present a vision of a new form of software systems, which we term \textbf{AI-Driven Self-Evolving Software}. At the outset, such software may provide no or only a basic set of functionality. Through continuous interaction with users, it can progressively enrich or modify its internal implementations, thereby evolving into a specialized software tailored to the user. By replacing costly human developers with AI, it can substantially reduce economic cost, and by enabling continuous self-evolution rather than proceeding through multiple independent stages, it can significantly decrease time overhead. This new form of software systems opens up the possibility of realizing genuine software automation.

\begin{figure}[!t]
    \centering
    \vspace{0.03in}
    \includegraphics[width=0.9\linewidth]{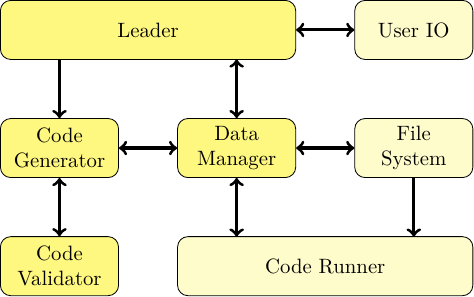}
    \vspace{-0.03in}
    \caption{Overall architecture of the proposed software system, consisting of four key modules: \textit{Leader}, \textit{Data Manager}, \textit{Code Generator}, and \textit{Code Validator}.}
    \label{fig:overview}
    \vspace{-0.17in}
\end{figure}

In this paper, we design a prototype of \textbf{AI-Driven Self-Evolving Software}, implemented through a multi-agent system~\cite{dorri2018multi}.
Although lightweight, we argue that this software system can be readily scaled to support more sophisticated functions, paving the way toward fully self-evolving software systems without human involvement.
We then evaluate this prototype through a series of representative case studies spanning diverse application scenarios, demonstrating the strong potential of \textbf{AI-Driven Self-Evolving Software}. Finally, we outline our future plans, summarizing directions for extending and deepening this line of work.

\section{AI-Driven Self-Evolving Software}

In this section, we present a prototype of \textbf{AI-Driven Self-Evolving Software}. We begin with an overview of the prototype (Section~\ref{sec:overview}) and then detail its key modules (Section~\ref{sec:leader}, \ref{sec:manager}, \ref{sec:generator}, and \ref{sec:validator}).

\subsection{Overview}
\label{sec:overview}

User requirements expressed in natural language often differ substantially from their corresponding software implementations. In the traditional software development lifecycle, this gap is addressed through a sequence of stages including requirement analysis, system design, implementation, and testing, which progressively transform requirements into executable software~\cite{ruparelia2010software, petersen2009waterfall, abrahamsson2017agile}. Inspired by this paradigm, we propose a multi-agent software system that incrementally refines itself to satisfy user requirements. As illustrated in Figure~\ref{fig:overview}, the prototype comprises four key modules, each realized through either an agent or an automated workflow: \textit{Leader}, \textit{Data Manager}, \textit{Code Generator}, and \textit{Code Validator}.

When user requirements change or new requirements emerge (\eg querying the weather in a given location), the system operates as follows.
\ding{182} The \textit{Leader} interprets user requirements and determines whether the current system already satisfies them. Based on this judgment, it decides either to invoke an existing functionality (\eg invoking the weather query functionality to obtain results) or to initiate self-evolving (\eg developing a new weather query functionality).
\ding{183} The \textit{Code Generator} implements new functionality in response to the identified needs (\eg writing the code to call a weather API).
\ding{184} Since AI-generated implementations are not guaranteed to be correct, the \textit{Code Validator} automatically performs extensive testing on newly developed functionalities to improve reliability.
\ding{185} Since self-evolving software requires continuous maintenance and organization, we introduce the \textit{Data Manager} to autonomously maintain its data, including source code and other related data.

\subsection{Leader}
\label{sec:leader}

The \textit{Leader} serves as the core of the AI-driven self-evolving software, acting both as the interface for user interaction and as the manager of the overall software system.

Specifically, the \textit{Leader}, implemented as an advanced agent, conducts a detailed analysis of user requirements and autonomously performs a series of actions to address diverse and changing needs, as detailed in the following:
\ding{182} \textbf{Analyze Existing Programs.} The \textit{Leader} can consult the \textit{Data Manager} to review or execute any program within the software system with appropriate arguments. It then evaluates the results to determine whether the user’s requirements are already satisfied or whether additional functionality is necessary.
\ding{183} \textbf{Request New Functionality.} If existing code does not fulfill the user’s requirements, the \textit{Leader} delegates the task to the \textit{Code Generator}, which either produces new code or modifies existing implementations.
\ding{184} \textbf{Satisfy User Requirements.} Once sufficient functionality has been implemented, the \textit{Leader} invokes the appropriate functionality and returns the results to the user.

\subsection{Data Manager}
\label{sec:manager}

\begin{figure}[!t]
    \centering
    \includegraphics[width=\linewidth]{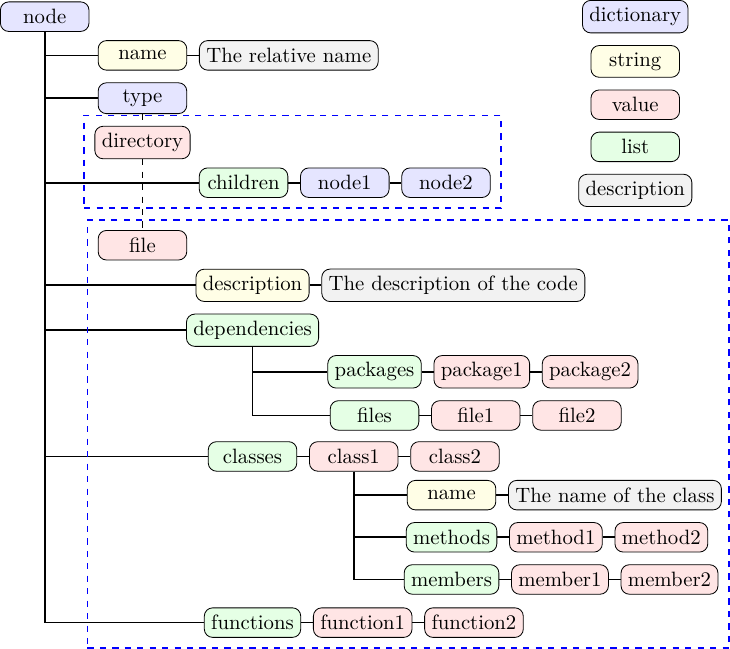}
    \caption{Hierarchical structure managed by the \textit{Data Manager}, with each node representing a directory or file and its metadata.}
    \label{fig:structure}
    \vspace{-0.18in}
\end{figure}

As the prototype is essentially a software system, it requires an autonomous module that continuously manages and organizes its data, including source code and other related resources. This presents a key challenge in the construction of the prototype.

\vspace{3pt}
\begin{boxK} \small \faIcon{exclamation-triangle} \textbf{\textbf{Key Challenge 1:}} 
How to effectively manage and organize growing software data in a reliable and scalable manner.
\end{boxK}

To address this challenge, we design the \textit{Data Manager}, which incorporates three key aspects:
\ding{182} \textbf{Storage of Data.} Traditional software systems typically rely on file systems for data storage. Following this practice, we employ a dedicated file system to store the data of the software. In anticipation of future AI-driven self-evolving software that may need to handle larger volumes of data, the prototype also provides interfaces for integration with database systems.
\ding{183} \textbf{Representation of Data.} As the scale of the software grows and given the inherent token limitations of agents, it is impractical to provide all data directly to other agents. To mitigate this issue, the \textit{Data Manager} employs a hierarchical representation of the data. Specifically, the data is organized into a tree structure that mirrors the directory hierarchy, as illustrated in Figure~\ref{fig:structure}. Each node represents either a directory or a file and is annotated with metadata describing its contents. This structured information is presented to other agents in JSON format, which they can efficiently interpret~\cite{agarwal2025think}.
\ding{184} \textbf{Functions.} Building on the aforementioned mechanisms for storage and representation, the \textit{Data Manager} effectively maintains the growing body of data, provides agents within the software with information about various resources, and is even capable of executing source code to obtain results.

\subsection{Code Generator}
\label{sec:generator}
The \textit{Code Generator} is an LLM-based agent responsible for implementing new functionality requested by the \textit{Leader}. It interacts with the \textit{Data Manager} to read existing files, install necessary dependencies into the environment, and generate or modify code as required. For each newly created or updated code artifact, it records the file path along with descriptive metadata (\eg purpose and usage instructions) in a predefined JSON format, enabling the \textit{Data Manager} to maintain and organize the software effectively.

\subsection{Code Validator}
\label{sec:validator}

The correctness of AI-generated code cannot be guaranteed~\cite{li2025correctness, fawareh2024investigates}. Following traditional software development practices~\cite{myers2004art}, we introduce another agent, the \textit{Code Validator}, to verify the generated code. After the \textit{Code Generator} completes its operations, the resulting code is evaluated against test cases produced by the \textit{Code Validator}. If the code passes these tests, it is merged into the file system through the \textit{Data Manager}. Otherwise, the \textit{Code Validator} returns error information to the \textit{Code Generator}, which then attempts to regenerate the code based on the feedback.
However, this further raises a concern that the correctness of AI-generated test cases cannot be guaranteed either.  

\vspace{3pt}
\begin{boxK} \small \faIcon{exclamation-triangle} \textbf{Key Challenge 2:} 
How to design tests that are as reliable as possible for evaluating the correctness of AI-generated code.
\end{boxK}

Inspired by \textsc{MBR-EXEC}~\cite{shi2022natural}, we employ a cross-validation approach to mitigate this challenge partially. For a given new functionality requested by the \textit{Leader}, the \textit{Code Generator} is prompted to independently produce $N$ candidate programs:
\begin{equation}
\mathcal{P}=\{p_{1},p_{2},\dots,p_{N}\},
\end{equation}
which are executed on a lightweight and input-only suite generated by the \textit{Code Validator}:
\begin{equation}
\mathcal{T}=\{t_{1},t_{2},\dots,t_{K}\}.
\end{equation}

We consider identical execution results (\ie not only identical outputs but also consistent effects on the overall software, including internal file systems and the runtime environment) as a strong indicator of semantic equivalence.
The hard mismatch loss between two candidates $p_{i}$ and $p_{j}$ is defined as follows:
\begin{equation}
\ell(p_{i},p_{j})=\max_{t\in\mathcal{T}} 1\bigl[p_{i}(t)\neq p_{j}(t)\bigr].
\end{equation}
Here, $p_{i}(t)$ denotes the execution result of program $p_{i}$ on input $t$. The empirical Bayes risk of program $p_{i}$ is then computed as
\begin{equation}
\mathrm{MBR\textrm-Risk}(p_{i})=\sum_{j=1}^{N}\ell(p_{i},p_{j}),
\end{equation}
where a lower risk reflects greater consistency in execution outcomes across the candidate pool. Additionally, we record the error count of each candidate as:
\begin{equation}
\mathrm{Err}(p_{i})=\sum_{k=1}^{K}1\bigl[p_{i}(t_{k})=\texttt{ERROR}\bigr].
\end{equation}
The final ranking rule is formulated as:
\begin{equation}
p^{*}=\arg\min_{p\in\mathcal{P}}\Bigl\langle\,\mathrm{MBR\textrm-Risk}(p),\;\mathrm{Err}(p)\;\Bigl\rangle,
\end{equation}
that is, by ascending order of MBR-Risk followed by ascending error count. Through cross-validation, the validator can select the highest-confidence code without any ground-truth outputs, relying only on raw textual programs and input examples, while also providing informative feedback for iterative repair.
\section{Case Design}

To provide a systematic evaluation of the prototype of AI-driven self-evolving software, we devised a set of representative cases. In this section, we present the design principles and specific choices of these cases.

\vspace{2.5pt}
\noindent \textbf{LLM-driven Agents.}  
All agents in the prototype (\ie \textit{Leader}, \textit{Code Generator}, and \textit{Code Validator}) are instantiated on top of the \texttt{o3}~\cite{o3}, a state-of-the-art reasoning LLM known for its strong capabilities in planning, code generation, and multi-step reasoning. The choice of \texttt{o3} ensures that each agent can reliably handle complex instructions, coordinate with other agents, and adapt to new user requirements.

\vspace{2.5pt}
\noindent \textbf{Case Scenarios.}  
We selected four representative scenarios (\ie user requirements) to evaluate the potential of the software and to uncover possible directions for future improvement. The selection was guided by the principle of covering diverse yet realistic software development scenarios, ensuring that the evaluation spans tasks with different requirements and levels of complexity.  
\ding{182} \textit{API Integration:} fetching weather forecasts from external services.  
\ding{183} \textit{Local Data Management:} creating and using a personal expense tracker.  
\ding{184} \textit{Web Resource Handling:} downloading repositories and files, and managing local assets.  
\ding{185} \textit{Text Processing:} implementing a Markdown-to-HTML converter.  

\vspace{2.5pt}
\noindent \textbf{Initial Setting.}  
At the beginning of all cases, the software system contained no predefined functionality. Each functionality was developed, validated, and integrated through the self-evolving process.

\vspace{2.5pt}
\noindent \textbf{Evaluation Strategy.}  
All evaluations were independently conducted by the three student authors, who systematically examined each case and verified the corresponding outcomes.
\section{Case Results}

In this section, we present the natural-language user requirements and the corresponding outcomes of several representative cases. Based on these cases, we further summarize a set of key findings.

\vspace{-4pt}
\subsection{API Integration}
\label{sec:case1}

\noindent \textbf{User Input.}
The user issued two prompts sequentially: 
(1) \textit{Please help me check the weather in Beijing for tomorrow and the day after tomorrow.}  
(2) \textit{I am currently in London, please help me check the weather for the next two days in London.}

\vspace{3pt}
\noindent \textbf{Result.}  

For the first prompt, the \textit{Leader} analyzed the requirement and determined that no suitable functionality was available in the existing system. It therefore delegated the task to the \textit{Code Generator}, which implemented a new Python component, weather\_forecast.py, to fetch weather data from a public API. The generated code was then verified by the \textit{Code Validator}, which executed multiple test cases and confirmed its correctness. After validation, the \textit{Data Manager} integrated the component into the local file system. The \textit{Leader} invoked this newly created functionality and returned the weather forecast in a natural-language response.  

\vspace{3pt}
\begin{boxK} \small \faIcon{pencil-alt} \textbf{Finding 1:} 
The software is capable of self-evolving by generating new functionality according to user requirements.
\end{boxK}

For the second prompt, the \textit{Leader} first examined the available functionality in the file system and identified that the existing weather\_forecast.py component already provided the required capability. It reused the component directly with updated arguments for London and responded to the user accordingly. 

\vspace{3pt}
\begin{boxK} \small \faIcon{pencil-alt} \textbf{Finding 2:} 
The software can effectively reuse previously generated functionality to efficiently address user requirements.
\end{boxK}

\subsection{Local Data Management}
\label{sec:case2}

\noindent \textbf{User Input.}  
The interaction between the user and the software consisted of three parts.  
(1) The user first requested the creation of a tool: \textit{I need a expense recorder to keep track of daily expenses, with fields including date, amount, category, and notes.}  
(2) Next, the user provided a sequence of individual expenses in natural language, such as \textit{I spent 58 yuan on dinner on September 1st, please help me keep a record.}, which required the software to parse and record entries dynamically.  
(3) Finally, the user asked for an analytical summary of the collected data: \textit{How much is expected to be spent in total? ... Create a table by category and summarize it for me.}

\vspace{3pt}
\noindent \textbf{Result.}  
For the first input, the system created a new component, expense\_recorder.py, which was designed to append expense entries to a local expenses.csv file. For each subsequent input, the \textit{Leader} parsed the natural-language description of an expense, extracted the relevant fields, and invoked the recorder to log the data in CSV format. When the user requested a summary, the \textit{Leader} accessed the recorded data and performed aggregation across categories, returning a formatted table that answered the user’s query. This demonstrates the system’s capability to not only generate tools but also reason over the data they produce.  

\vspace{3pt}
\begin{boxK} \small \faIcon{pencil-alt} \textbf{Finding 3:} 
The software is able to handle multi-step user requirements, formulated as long sequences of related tasks.
\end{boxK}

\subsection{Web Resource Handling}
\label{sec:case3}

\noindent \textbf{User Input.}  
The interaction was designed to cover three representative tasks that reflect common user requirements:
(1) downloading a GitHub repository, (2) downloading a PDF article, and (3) deleting the downloaded file.

\vspace{3pt}
\noindent \textbf{Result.}  
For all three tasks, the software successfully completed the operations, including downloading web resources and subsequently deleting them from the local system. Notably, in this case the \textit{Code Generator} produced scripts that did not provide any outputs. To address this, the \textit{Code Validator} verified correctness by inspecting the state of the file system (\eg ensuring that multiple candidate programs consistently deleted the same file). This mechanism ensured reliable validation even in the absence of direct program outputs.

\vspace{3pt}
\begin{boxK} \small \faIcon{pencil-alt} \textbf{Finding 4:} 
The software can reliably validate diverse functionality by reasoning about external environmental states.
\end{boxK}

\subsection{Text Processing}
\label{sec:case4}

\noindent \textbf{User Input.}  
The user requested the conversion of a Markdown file located in the file system into an HTML document, for example: \textit{Please convert the file at ./docs/test.md into HTML format.}

\vspace{3pt}
\noindent \textbf{Result.}  
The \textit{Code Generator} produced a script, converter.py, which was integrated into the system by the \textit{Data Manager}. When executed by the \textit{Leader}, the script successfully transformed the specified Markdown file into output.html. Inspection of the generated HTML confirmed that the structural elements of the source Markdown document, such as headers, lists, and code blocks, were accurately preserved in the output.

\section{Future Plans}

The prototype presented in this paper demonstrates the feasibility of \textbf{AI-Driven Self-Evolving Software}. However, it represents only a preliminary step toward a broader vision. We outline several promising directions for future exploration.

\textbf{Scaling to Complex Scenarios.} The current prototype focuses on relatively lightweight scenarios. A key direction is to extend the self-evolving software system to support more complex and large-scale applications that require integration across multiple functionalities and external APIs. This will test the scalability of the basic architecture established in the current prototype.

\textbf{Establishing Benchmarks for Evaluation.} At present, there is no systematic benchmark to evaluate the capabilities of our proposed AI-driven self-evolving software system, particularly its ability to evolve without human developers. Future work will focus on developing benchmarks that encompass diverse self-evolution scenarios, thereby enabling rigorous and comparative evaluations of AI-driven self-evolving software systems.

\textbf{Enhancing Reliability and Trustworthiness.} Although our design incorporates validation through cross-execution consistency, challenges remain in ensuring the correctness of AI-generated code. In the near future, we will investigate stronger verification strategies, including formal methods and runtime monitoring, to improve trustworthiness in safety-critical contexts. 

\textbf{Toward Full Self-Evolution.} Our long-term goal is to enable software systems to sustain continuous self-evolution, thereby moving closer to genuine software automation. Instead of merely responding to explicit user requirements, future software systems should proactively identify limitations, propose new capabilities, and reorganize their data over time. Realizing such autonomy calls for advances in adaptive planning and long-term memory, forming a long-term research agenda at the intersection of AI and software engineering.

\section{Conclusion}
This paper presents a vision for \textbf{AI-Driven Self-Evolving Software}, a new form of software systems that can autonomously enrich and adapt their functionality through continuous interaction with users. We developed a lightweight prototype to illustrate this idea and evaluated it across representative cases. These results highlight both the feasibility and potential of self-evolving software. Although still at an early stage, this line of work paves the way toward scalable self-evolving software systems, offering a promising step toward realizing genuine software automation.

\bibliographystyle{ACM-Reference-Format}
\bibliography{main}

\end{document}